\documentclass{jltp}

\usepackage{epsfig}
\newcommand{\be}{\begin{equation}}
\newcommand{\ee}{\end{equation}}
\newcommand{\bea}{\begin{eqnarray}}
\newcommand{\eea}{\end{eqnarray}}
\newcommand{\de}{{\rm d}}
\newcommand{\ex}[1]{{\rm e}^{#1}}
\newcommand{\dg}{^{\dagger}}
\newcommand{\eps}{\epsilon}
\newcommand{\nn}{\nonumber \\}
\newcommand{\pick}[2]{#1} % #1 for color, #2 for black and white

\title{Stability Analysis of the Hubbard Model}

\author{I. Grote, E. K\"ording, and F. Wegner}

\address{Institut f\"ur Theoretische Physik,
Ruprecht-Karls-Universit\"at\\  Philosophenweg 19, D-69120 Heidelberg}

\runninghead{I. Grote, E. K\"ording, and F. Wegner}{Stability Analysis
of Hubbard Model}

\begin{document}

\maketitle

\begin{abstract}

An effective Hartree-Fock-Bogoliubov-type interaction is calculated for the
Hubbard model in second order in the coupling by means of flow equations.
A stability analysis is performed in order to obtain the transition into
various possible phases. 

We find, that the second order contribution weakens the tendency for the
antiferromagnetic transition. Apart from a possible antiferromagnetic
transition the $d$-wave Pomeranchuk instability recently reported by Halboth
and Metzner is usually the strongest instability. A newly found instability
is of $p$-wave character and yields band-splitting. In the BCS-channel one
obtains the strongest contribution for $d_{x^2-y^2}$-waves. Other types of
instabilities of comparable strength are also reported.


\end{abstract}

\section{INTRODUCTION}

The Hubbard model is commonly considered a model which contains
essential features of the Cuprate layers of high-temperature
superconductors.\cite{Scalapino}

In this paper we analyze the Hubbard model by means of flow
equa\-tions\cite{Wegner}. The basic idea is to eliminate off-diagonal
matrix-elements of the interaction in order to obtain a diagonal or a
block-diagonal Hamiltonian. This is done by a continuous unitary
transformation of the Hamiltonian as a function of the flow-parameter $l$,
\be
\frac{\de H(l)}{\de l} = [\eta(l),H(l)]. \label{flow}
\ee

In the original paper\cite{Wegner} the Hamiltonian $H$ was
decomposed into two parts, the diagonal part $H_{\rm d}$ and the off-diagonal
part $H_{\rm r}$ and the generator $\eta$ of the flow equation was chosen so
that $H_{\rm r}$ was eliminated. This could be done for example by choosing
\be
\eta=[H_{\rm d},H]. \label{eta}
\ee
In the case of interacting electrons the diagonal terms in (\ref{eta})
were those conserving the number of quasi-particles, i.e. electrons above and
holes below the Fermi-sea. Thus the contributions creating and
annihilating quasi-particles across the Fermi-surface were eliminated.
This scheme was useful at zero temperature. However, at finite temperatures
the Fermi-surface is no longer sharp. Then this scheme is no longer useful.

Instead we will choose a different approach. We rewrite eq.(\ref{eta})
\be
\eta=[H,H_{\rm r}],
\ee
and in contrast to our previous approach we consider no longer terms to be
either diagonal or off-diagonal, but we introduce a continuous quantity $r>0$
associated to the terms in the Hamiltonian $H_{\rm r}$ which determines how
{\em urgently} we wish to {\em eliminate} the corresponding term. $r=0$ means,
that we keep the term. The larger $r$ the more urgently we eliminate it. We
call $r$ the {\rm elimination factor}. In our previous scheme the terms in the
diagonal part $H_{\rm d}$ have $r=0$, the terms in the off-diagonal part
$H_{\rm r}$ have $r=1$. From now on $r$ may be continuous.

We will use this scheme in order to transform the Hamiltonian into the form of
a molecular-field type Hamiltonian, that is into a form which depends only on
biquadratic terms $c\dg_k c_k$, $c\dg_k c\dg_{-k}$, and $c_{-k}c_k$, which
yields a Hartree-Fock-Bogoliubov-type interaction. In order to
accomodate antiferromagnetism, too, we will keep terms $c\dg_{k+q_0} c_k$, and
also $c\dg_k c\dg_{q_0-k}$, and $c_{q_0-k}c_k$, where
$q_0=(\frac{\pi}a,\frac{\pi}a)$ is the staggered wave-vector for lattice
spacing $a$. This yields an effective interaction which contains the
interaction terms described in eqs. (\ref{HB},\ref{HHF},\ref{HY}, and
\ref{HAC}). In this way we eliminate the fluctuations around the
molecular-field type behavior. The calculation of this effective interaction
is performed in second-order in the Hubbard-coupling $U$.

This scheme differs from the renormalization-group scheme described by
Shankar\cite{Shankar} and applied by Zanchi and Schulz\cite{Schulz}, by
Salmhofer\cite{Salmhofer} and by Halboth and Metz\-ner\cite{Metzner}, and by
Honerkamp et al.\cite{Honerkamp} to the Hubbard model in so far, as they
eliminate the degrees of freedom outside a shell enclosing the Fermi-surface.
In our scheme however, first, the interactions connecting states far apart in
energy, irrespective of their distance from the Fermi-level, are eliminated.
This is in the same spirit as the similarity renormalization by G{\l}azek and
Wilson.\cite{Glazek} Since we transform the interaction we always deal with an
effective interaction and not with truncated and partially integrated
correlations. This implies, that the interaction remains finite at
phase-transitions.

Starting from the Hartree-Fock-Bogoliubov interaction we perform a stability
analysis, that is we investigate, at which critical $(U/t)_c$ becomes the
system unstable against any symmetry breaking contributions. It has long been
known, that at half-filling the system becomes antiferromagnetic, and during
recent years many calculations have shown, that the Hubbard model shows an
attractive interaction for $d_{x^2-y^2}$-wave pairing. Apart from the
above mentioned\cite{Schulz,Metzner} We refer to the FLEX-approach by
Bickers, Scalapino, and White\cite{BickersI,BickersII}, the review by
Scalapino\cite{Scalapino} and calculations by Hanke et al.\cite{Hanke}.
We find that often the strongest instability (apart from
antiferromagnetism) is a $d$-wave Pomeranchuk-instability recently reported by
Halboth and Metz\-ner\cite{MetznerII} and a band-splitting instability of
$p$-wave character. The $d$-wave flux-phase instability
(Kotliar\cite{Kotliar}, Affleck and Marston\cite{Affleck}, Chakravarty et
al.\cite{Chakravarty}) appears degenerate with the superconductiving
instability at half filling, and above at doping.

In section 2 we derive a general expression for the generator $\eta$ from a
gradient procedure and introduce the elimination factor $r$. In section 3 the
second-order contribution to the effective electron-electron interaction is
derived. In section 4 the expressions for the free energy of the Hubbard model
are given. In section 5 the numerical results for the phase-instabilities are
presented.

\section{GENERATOR $\eta$ FOR FLOW EQUATIONS FROM A GRADIENT PROCEDURE}

In order to determine the generator $\eta$ in the flow equation (\ref{flow}) we
introduce a quadratic form of the Hamiltonian
\be
G(H)=\frac 12 \sum g_{ij,kl} H_{ji} H_{lk}.
\ee
This form is chosen in such a way that $G$ becomes a minimum by means
of the flow equations. Without restriction of the general procedure we
require the symmetry
\be
g_{ij,kl}=g_{kl,ij}.
\ee
$G$ should be real. Since $H$ is hermitean we have
\be
G^*(H) = \frac 12 \sum g^*_{ij,kl} H_{ij} H_{kl}.
\ee
Thus $G^*=G$ is obtained for
\be
g_{ij,kl}=g^*_{ji,lk}.
\ee
We consider now the variation of $G$ under the flow
\bea
\frac{\de G}{\de l}&=&\sum g_{ij,kl} (\eta_{jm} H_{mi} - H_{jm} \eta_{mi})
H_{lk} \nn &=& \sum \eta_{ji} (g_{mi,kl} H_{im} H_{lk} - g_{im,kl} H_{mj}
H_{lk}). \eea
Since $dG/dl$ should be (semi-)negative, we choose
\be
\eta_{ji} = -(g_{mj,kl} H_{im} H_{lk} -g_{im,kl} H_{mj} H_{lk})^*
= -g_{jm,kl} H_{mi} H_{kl} + g_{mi,lk} H_{jm} H_{kl},
\ee
which can be rewritten
\be
\eta = [H,H_{\rm r}]
\ee
with
\be
H_{{\rm r},ji} = g_{ij,lk} H_{kl}.
\ee
$G$ should decrease under the flow. Therefore one chooses a quadratic form,
which vanishes for the desired diagonal form, but whose other eigenvalues
are positive. For this purpose one may use a double commutator with a hermitean
operator $v$,
\be
H^r=[v,[v,H]].
\ee
With this choice those contributions to $H$ are considered diagonal, which 
commute with $v$. If for example $v$ is the operator of the quasi-numbers, 
then a matrix element $H_{ij}$ which connects states with $v_i$ and $v_j$ 
quasi-particles contributes $(v_i-v_j)^2 H_{ij} H_{ji}$ to $G(H)$.
For a general basis one has
\be
g_{ji,kl}=(v^2)_{jk} \delta_{li} - 2v_{jk} v_{li} +\delta_{jk} (v^2)_{li}.
\ee
This matrix $g$ is symmetric and hermitean as required.

Instead of using one double-commutator one may use a sum of such contributions.
This can be used for the investigation of systems at finite
temperatures where due to the disappearance of a sharp Fermi-surface the
concept of particles above and holes below the Fermi-surface is no longer
valid. We describe the fermion system in terms of creation and annihilation 
operators $c\dg$ and $c$ and introduce the operator $v$ as a one particle
operator \be
v=\sum v_k c\dg_k c_k.
\ee
Then we obtain the contributions of $H_{\rm r}$ by multiplying the terms \\
$H_{k_1,k_2,...k'_1,k'_2,...} c\dg_{k_1} c\dg_{k_2} ... c_{k'_1} c_{k'_2} ...$
of $H$ by the elimination function 
\be
r_{k_1,k_2,...,k'_1,k'_2,...}=(v_{k_1}+v_{k_2}+...-v_{k'_1}-v_{k'_2}-...)^2.
\ee
It indicates how  strongly we wish to eliminate this term. 
The larger it is the more we consider it to be non-diagonal.
If we now choose several of these contributions with functions
\be
v^{(\alpha)}_k = (1-p^2)^{1/4} \frac{p^{\alpha/2} H_{\alpha}(k/k_0)}
{2^{\alpha/2}\sqrt{\alpha!}} e^{-pk^2/((1+p)k_0^2)} \label{vhermite}
\ee
with hermite polynomials $H_{\alpha}(x)$ and sum over all $\alpha$ (in $d$
dimensions one has to introduce products of $d$ such functions
\be
v^{\alpha_1,...\alpha_d}_{\bf k} = v^{\alpha_1}_{k_1} ... v^{\alpha_d}_{k_d}
\ee
and to sum over all of them) then one obtains for the term \\
$H_{k_1,k_2,...k'_1,k'_2,...} c\dg_{k_1} c\dg_{k_2} ... c_{k'_1} c_{k'_2} ...$
the elimination factor
\bea
r_{k_1,k_2,...k'_1,k'_2,...}&=&
\sum_{i,j} r(k_j,k_j) -2\sum_{i,j} r(k_i,k'_j) +\sum_{i,j} r(k'_i,k'_j), 
\label{elfkt} \\
r(k,k')&=& \exp(-\frac p{1-p^2}(k-k')^2/k_0^2).
\eea
Use is made of the generating function (19.12.14) in \cite{Erdelyi}.
Here $p$ and $k_0$ are appropriately chosen parameters. This function $r$ 
has the following nice properties: 
If the momentum $k$ of a creation operator and the momentum $k'$ of an 
annihilation operator are equal, then the function $r$ is the same if we remove
these two operators. Therefore multiplication of a term with the number
operator $c\dg_k c_k$ will not change $r$. Consequently if all the momenta
$k$ of $c\dg$ and $k'$ of $c$ are pairwise equal, then the factor $r$ 
vanishes. Indeed these matrix-elements to the hamiltonian are considered to
be diagonal. If the differences between all momenta are large in comparison to
$k_0$, then only the contributions for $i=j$ are left, so that $r$ approaches
the sum of the number of creation and annihilation operators. If the number
of creation operators is different from the number of annihilation operators,
then for nearly equal momenta the factor $r$ becomes the square of the
difference of the number of creation and annihilation operators.

If a system is to undergo a superconducting transition then one should keep
pairs of creation operators with momenta $k$ and $-k$, similarly for
annihilation operators. In this case it is more appropriate to sum only
over the contributions with odd $\alpha$ in (\ref{vhermite}). In $d$
dimensions one keeps those with odd $\alpha_1 + \alpha_2 + ...\alpha_d$.
For these odd $\alpha$ the contribution $v_k+v_{-k}$ vanishes, so that
multiplication by $c\dg_k c\dg_{-k}$ or $c_kc_{-k}$ does not alter
the factor $r$. If a matrix-element consists only of such factors and
factors $c\dg_k c_k$, then its $r$ vanishes. With this choice one obtains the
elimination factor (\ref{elfkt}) with 
\bea
r(k,k')= \frac 12 \left(\exp(-\frac p{1-p^2}(k-k')^2/k_0^2)
-\exp(-\frac p{1-p^2}(k+k')^2/k_0^2)\right).
\eea

It should be mentioned, that $G$ need not necessarily be a quadratic form
in the Hamiltonian. For a general differentiable $G$ one obtains
\be
\frac{\de G}{\de l} = \sum \frac{\partial G}{\partial H_{ij}} 
(\eta_{ik} H_{kj} -H_{ik} \eta_{kj})
=\sum \eta_{ji} (H_{ik} \frac{\partial G}{\partial H_{jk}} - H_{kj} 
\frac{\partial G}{\partial H_{ki}}).
\ee
Choosing now 
\be
\eta_{ji}=-(H_{ik} \frac{\partial G}{dH_{jk}} 
- H_{kj} \frac{\partial G}{\partial H_{ki}})^*
\ee
we obtain
\be
\eta=[H,R]
\ee
with
\be
R_{ji} = \frac{\partial G}{\partial H_{ij}}.
\ee
The choice 
\be
G=-\sum k H_{kk}
\ee
orders the states with increasing energies\cite{Mielkeprivat}. One obtains
\be
R_{ji} = -j \delta_{ji}, \quad \eta_{ji}=(i-j) H_{ji}.
\ee
Other choices of $G$ linear in $H$ are
the  introduction of $\eta_{ji} = {\rm sign}(i-j) H_{ji}$ which also orders
the states in increasing order\cite{Mielke,SteinLipkin} or choosing $R$ being
the double-occupancy in the Hubbard model\cite{SteinHubbard}.

Note that the introduction of $G$ gives a straighforward way to obtain an
$\eta$ which minimizes $G$. However it is not the only way to do this.
A counterexample is the modification used for the spin-boson model\cite{KMN}.

\section{SECOND-ORDER PERTURBATON THEORY}

We now use this scheme in order to determine the flow equations
\be
\frac{\de H(l)}{\de l}=[\eta(l),H(l)]=[[H,V_r],H]=[[T+V,V_r],T+V]
\ee
up to second-order in the interaction $V$
\bea
H&=&T+V \\
T&=&\sum_{q,s} \eps_q :c\dg_{qs} c_{qs}: \\
V&=&\frac 1{2\Omega} \sum_{k_1,q_1,s_1,k_2,q_2,s_2} V(k_1,k_2,q_1,q_2)
:c\dg_{k_1s_1} c_{q_1s_1} c\dg_{k_2s_2} c_{q_2s_2}:.
\eea
where $\Omega$ is the volume of the system. In first order
\be
\frac{\de V_1(l)}{\de l}=[[T,V_{\rm 1r}],T]
\ee
one obtains
\be
\frac{\de V_1(k_1,k_2,q_1,q_2;l)}{\de l}=
-\Delta\eps(k_1,k_2,q_1,q_2)^2 r(k_1,k_2,q_1,q_2) V_1(k_1,k_2,q_1,q_2;l)
\ee
with
\be
\Delta\eps(k_1,k_2,q_1,q_2)=\eps_{k_1}+\eps_{k_2}-\eps_{q_1}-\eps_{q_2}
\ee
and
\be
V_1(k_1,k_2,q_1,q_2;l)=
\ex{-\Delta\eps(k_1,k_2,q_1,q_2)^2 r(k_1,k_2,q_1,q_2)l} V(k_1,k_2,q_1,q_2).
\ee
Thus the decay of the off-diagonal terms does not only contain the factor 
$(\Delta\eps)^2$ in the exponential, but also the elimination factor $r$.
The one-particle contribution $T$ does not change in first order in $V$. We
will neglect the second-order contribution to $T$.

In second order in $V$ we have
\be
\frac{\de V_2(l)}{\de l}-[[T,V_{\rm 2r}],T]=[[T,V_{\rm 1r}],V_1]+[[V_1,V_{\rm
1r}],T] \label{zweiteOrd}
\ee
{From} this equation we obtain
\bea
\big(\frac{\de}{\de l} +\Delta\eps(k_1,k_2,q_1,q_2)^2 
r(k_1,k_2,q_1,q_2)\big) V_2(k_1,k_2,q_1,q_2;l) = \nn
\frac 1{\Omega}\sum_{p_1,p_2} f(k_1,k_2,p_1,p_2;p_1,p_2,q_1,q_2;l)
(1-n_{p_1}-n_{p_2}) \nn
+\frac 2{\Omega}\sum_{p_1,p_2} f(k_1,p_1,q_1,p_2;k_2,p_2,q_2,p_1;l)
(n_{p_1}-n_{p_2}) \nn
-\frac 1{\Omega}\sum_{p_1,p_2} f(k_1,p_1,p_2,q_1;k_2,p_2,q_2,p_1;l)
(n_{p_1}-n_{p_2}) \nn
-\frac 1{\Omega}\sum_{p_1,p_2} f(k_1,p_1,q_1,p_2;k_2,p_2,p_1,q_2;l)
(n_{p_1}-n_{p_2}) \nn
-\frac 1{\Omega}\sum_{p_1,p_2} f(k_1,p_1,p_2,q_2;k_2,p_2,p_1,q_1;l)
(n_{p_1}-n_{p_2})
\eea
with
\bea
f(\{a\};\{b\};l)=V_1(\{a\};l)V_1(\{b\};l) \nn \times
\left(r(\{a\})(2\Delta\eps(\{a\})+\Delta\eps(\{b\}))
-r(\{b\})(\Delta\eps(\{a\})+2\Delta\eps(\{b\}))\right).
\eea
For those momenta which obey $r(k_1,k_2,q_1,q_2)=0$ we obtain
\bea
V_2(k_1,k_2,q_1,q_2;\infty)=\nn
\frac 1{\Omega}\sum_{p_1,p_2} F(k_1,k_2,p_1,p_2;p_1,p_2,q_1,q_2)
(1-n_{p_1}-n_{p_2}) \nn
+\frac 2{\Omega}\sum_{p_1,p_2} F(k_1,p_1,q_1,p_2;k_2,p_2,q_2,p_1)
(n_{p_1}-n_{p_2}) \nn
-\frac 1{\Omega}\sum_{p_1,p_2} F(k_1,p_1,p_2,q_1;k_2,p_2,q_2,p_1)
(n_{p_1}-n_{p_2}) \nn
-\frac 1{\Omega}\sum_{p_1,p_2} F(k_1,p_1,q_1,p_2;k_2,p_2,p_1,q_2)
(n_{p_1}-n_{p_2}) \nn
-\frac 1{\Omega}\sum_{p_1,p_2} F(k_1,p_1,p_2,q_2;k_2,p_2,p_1,q_1)
(n_{p_1}-n_{p_2}) \label{V_2}
\eea
with
\bea
F(\{a\};\{b\})=V(\{a\})V(\{b\}) \nn \times
\frac{r(\{a\})(2\Delta\eps(\{a\})+\Delta\eps(\{b\}))
-r(\{b\})(\Delta\eps(\{a\})+2\Delta\eps(\{b\}))}
{r(\{a\})\Delta\eps(\{a\})^2+r(\{b\})\Delta\eps(\{b\})^2}.
\eea
Since the initial condition $V(k_1,k_2,q_1,q_2)=U\delta_{k_1+k_2,q_1+q_2}$
of the Hubbard model as well as $r$ and $\Delta\eps$ are
invariant upon exchange of the last two arguments and independently
against exchange of the first two arguments, the same holds for $V_1$. 
Consequently $F$ obeys the symmetry relations
\bea
F(a_1,a_2,a_3,a_4;\{b\})=F(a_2,a_1,a_3,a_4;\{b\})=F(a_1,a_2,a_4,a_3;\{b\}) \\
F(\{a\};b_1,b_2,b_3,b_4)=F(\{a\};b_2,b_1,b_3,b_4)=F(\{a\};b_1,b_2,b_4,b_3).
\eea
Then the second, third and fourth term in the expression (\ref{V_2}) for $V_2$
cancel.

\subsection{Effective Interaction in various Channels}

We wish now to retain all terms in the interaction which seem to be important
for phase-transitions in the Hubbard model. In particular we are interested to
keep those relevant for antiferromagnetism and superconductivity that is terms
of the form $c\dg_{k+q_0} c_k c\dg_q c_{q+q_0}$ and 
$c\dg_k c\dg_{-k} c_{-q} c_q$. 
We use the elimination factor $r$ in the form
\be
r(k_1,k_2,q_1,q_2)=\sum_{\alpha} 
\left(v^{\alpha}(k_1)+v^{\alpha}(k_2)-v^{\alpha}(q_1)-v^{\alpha}(q_2)\right)^2.
\ee
In order to keep the interaction terms relevant for antiferromagnetism we
choose
\be
v^{\alpha}(q)=v^{\alpha}(q+q_0). \label{v1}
\ee
Further, in order to keep the pair-interaction we require
\be
v^{\alpha}(q)=-v^{\alpha}(-q). \label{v2}
\ee
With this choice the contributions listed in the first column of table 1 will
be kept. Obviously, $V_{\rm B}$ contains the effective interaction between
pairs of electrons of total momentum 0, $V_{\rm A}$ contains the effective
antiferromagnetic interaction with staggered wave-vector $q_0$. $V_{\rm C}$
and $V_{\rm A}$ contain also charge waves. The contributions $V_{\rm H}$ and
$V_{\rm F}$ contain the contributions kept in the Fermi-liquid picture. With
the choice (\ref{v1}) and (\ref{v2}) also the terms $V_{\rm Y}$ survive, which
describe the pair-interaction of electron pairs with total momentum $q_0$.

With this choice of the elimination function $r$ it turns out, that in all
cases in the table we obtain $r(\{a\})=r(\{b\})$. Then these factors cancel
altogether and the expression for $F(\{a\};\{b\})$ reduces to
\be
F(\{a\};\{b\})=V(\{a\})V(\{b\})
\frac{\Delta\eps(\{a\})-\Delta\eps(\{b\})}
{\Delta\eps(\{a\})^2+\Delta\eps(\{b\})^2}.
\ee

\bea
\renewcommand{\sc}[1]{\scriptstyle{#1}}
\begin{array}{|c|c|cc|c|}
\hline 
\sc V & \sc W & \sc{\{a\}} & \sc{\{b\}} \\ \hline
\sc{V_{\rm 2B}(k,q)=V_2(k,-k,q,-q)} & \sc 0 &\sc{(k,-k,p_1,p_2)} &
\sc{(p_1,p_2,q,-q)} \\
\sc{V_{\rm 2B}=W_{\rm B}} & \sc{W_{\rm B}} & \sc{(k,p_1,p_2,-q)} &
\sc{(-k,p_2,p_1,q)} \\ \hline
\sc{V_{\rm 2H}(k,q)=V_2(k,q,k,q)} & \sc{W_{\rm F}} &
\sc{(k,q,p_1,p_2)} & \sc{(p_1,p_2,k,q)} \\
\sc{V_{\rm 2H}=W_{\rm H}+W_{\rm F}}  & \sc{W_{\rm H}} &
\sc{(k,p_1,p_2,q)} & \sc{(q,p_2,p_1,k)} \\ \hline
\sc{V_{\rm 2F}(k,q)=V_2(k,q,q,k)} & \sc{W_{\rm F}} & \sc{(k,q,p_1,p_2)} &
\sc{(p_1,p_2,q,k)} \\
\sc{V_{\rm 2F}=W_{\rm F}}  & \sc 0 & \sc{(k,p_1,p_2,k)} & \sc{(q,p_2,p_1,q)} \\
\hline 
\sc{V_{\rm 2A}(k,q)=V_2(k,q+q_0,q,k+q_0)} & \sc{W_{\rm A}} &
\sc{(k,q+q_0,p_1,p_2)} & \sc{(p_1,p_2,q,k+q_0)} \\
\sc{V_{\rm 2A}=W_{\rm A}}  & \sc 0 & \sc{(k,p_1,p_2,k+q_0)} &
\sc{(q+q_0,p_2,p_1,q)} \\ \hline
\sc{V_{\rm 2C}(k,q)=V_2(k,q+q_0,k+q_0,q)} & \sc{W_{\rm A}} &
\sc{(k,q+q_0,p_1,p_2)} & \sc{(p_1,p_2,k+q_0,q)} \\ 
\sc{V_{\rm 2C}=W_{\rm C}+W_{\rm A}} & \sc{W_{\rm C}} & \sc{(k,p_1,p_2,q)} &
\sc{(q+q_0,p_2,p_1,k+q_0)} \\ \hline
\sc{V_{\rm 2Y}(k,q)=V_2(k,q_0-k,q,q_0-q)} & \sc 0 & \sc{(k,q_0-k,p_1,p_2)} &
\sc{(p_1,p_2,q,q_0-q)} \\
\sc{V_{\rm 2Y}=W_{\rm Y}}  & \sc{W_{\rm Y}} & \sc{(k,p_1,p_2,q_0-q)} &
\sc{(q_0-k,p_2,p_1,q)} \\ \hline
\end{array} \nonumber
\eea
Table 1. Matrix-elements in second order perturbation theory. In the first line the
data of the first term in (\ref{V_2}) are given, in the second line those of
the fifth term. In a number of cases $r=0$. These terms do not contribute.
They are indicated by $W=0$ in the table.\bigskip

This has the big advantage that we need not choose the functions $v$
explicitely, which prevents us from any arbitrariness in the evaluation.

The terms in the upper lines can be rewritten
\bea
W_{\rm F,A} &=&
\frac 1{\Omega}\sum_{p_1,p_2} F(k_1,k_2,p_1,p_2;p_1,p_2,q_1,q_2)
(1-n_{p_1}-n_{p_2}) \nn
&=& - \frac{U^2}{\Omega} \sum \frac{(1-n_{p_1}-n_{p_2})(\eps_{p_1}+\eps_{p_2}
-\frac{\eps^a+\eps^b}2)}{(\eps_{p_1}+\eps_{p_2}-\frac{\eps^a+\eps^b}2)^2
+(\frac{\eps^a-\eps^b}2)^2} \delta_{p_1+p_2,s} \label{FA}
\eea
using
\bea
\Delta\eps(\{a\}) &=& \eps^a - \eps_{p_1} - \eps_{p_2} \\
\Delta\eps(\{b\}) &=& \eps_{p_2} + \eps_{p_1} - \eps^b
\eea
and
\bea
\begin{array}{|c|cc|c|} \hline
 & \eps^a & \eps^b & s \\ \hline
W_{\rm F} & \eps_k+\eps_q & \eps_k+\eps_q & k+q \\
W_{\rm A} & \eps_k+\eps_{q+q_0} & \eps_q+\eps_{k+q_0} & k+q+q_0 \\ \hline 
\end{array} \nonumber
\eea
\centerline{Table 2. $\eps^a$, $\eps^b$, and $s$ in eq. (\ref{FA}).}
\medskip

The terms in the lower line can be written
\bea
W_{\rm B,H,C,Y}&=& 
-\frac 1{\Omega}\sum_{p_1,p_2} F(k_1,p_1,p_2,q_2;k_2,p_2,p_1,q_1)
(n_{p_1}-n_{p_2}) \nn 
&=& -\frac{U^2}{\Omega} \sum
\frac{(n_{p_1}-n_{p_2})(\eps_{p_1}-\eps_{p_2}
+\frac{\eps^a+\eps^b}2)}{(\eps_{p_1}-\eps_{p_2}+\frac{\eps^a+\eps^b}2)^2
+(\frac{\eps^a-\eps^b}2)^2} \delta_{p_2-p_1,s} \label{BHCY} 
\eea
with
\bea
\Delta\eps(\{a\}) &=& \eps^a + \eps_{p_1} - \eps_{p_2} \\
\Delta\eps(\{b\}) &=& \eps_{p_2} - \eps_{p_1} - \eps^b
\eea
and
\bea
\begin{array}{|c|cc|c|} \hline
 & \eps^a & \eps^b & s \\ \hline
W_{\rm B} & \eps_k-\eps_q & \eps_q-\eps_k & k+q \\
W_{\rm H} & \eps_k-\eps_q & \eps_k-\eps_q & k-q \\
W_{\rm C} & \eps_k-\eps_q & \eps_{k+q_0}-\eps_{q+q_0} & k-q \\
W_{\rm Y} & \eps_k-\eps_{q_0-q} & \eps_q-\eps_{q_0-k} & k+q-q_0 \\ \hline
\end{array} \nonumber
\eea
\centerline{Table 3. $\eps^a$, $\eps^b$, and $s$ in eq. (\ref{BHCY}).}
\medskip

Using $\eps_{-p}=\eps_p$ both terms can be written in the form
\be
- \frac{U^2}{\Omega} \sum \frac{(\hat n_{p_2}-n_{p_1})
(\eps_{p_2}+\hat \eps_{p_1}-\frac{\eps^a+\eps^b}2)}
{(\eps_{p_2}+\hat \eps_{p_1}-\frac{\eps^a+\eps^b}2)^2
+(\frac{\eps^a-\eps^b}2)^2} \delta_{p_1+p_2,s}. \label{sum}
\ee
For the first term $W_{\rm F,A}$ we have
\be
\hat n_{p_2} = 1-n_{p_2}, \quad \hat \eps_{p_1}=\eps_{p_1},
\ee
whereas for the last term $W_{\rm B,H,C,Y}$ one puts
\be
\hat n_{p_2} = n_{p_2}, \quad \hat \eps_{p_1}=-\eps_{p_1}.
\ee

\subsection{Isotropic Model}

Before going for the Hubbard Model let us consider an isotropic electron gas
without any lattice with one particle energy and interaction
\be
\eps_k=\frac{k^2}{2m}, \quad 
V(k_1,k_2,q_1,q_2) = U \delta_{k_1+k_2,q_1+q_2}.
\ee
The numerical calculations were performed by K\"ording and are
taken partly from\cite{Koerding}. They yield for the three-dimensional system
at zero-temperature with dimensionless quantities $k_{\rm F}=10$, $m=1$, the
following $p$-, $d$-, and $f$- wave-couplings $W_{\rm B}$ at the Fermi-surface
$-0.1174 U^2$, $-0.01808 U^2$, and $-0.006241 U^2$,  resp. Thus the couplings
decay rapidly with increasing angular momentum, but they are attractive.

In two dimensions the numerical calculation yields as a function of
temperature \\ 
\begin{center}
\begin{tabular}{|c | r r |}
\hline
$1/T$ & $p$-wave & $d$-wave \\ \hline
 & \multicolumn{2}{c|}{coefficient of} \\
 & $U^2 \cos(\theta)$ & $U^2 \cos(2\theta)$ \\ \hline
1 & $-2.18\cdot 10^{-3}$ & $-2.02\cdot 10^{-3}$ \\
2 & $-1.10\cdot 10^{-3}$ & $-1.06\cdot 10^{-3}$ \\
3 & $-7.32\cdot 10^{-4}$ & $-7.14\cdot 10^{-4}$ \\
4 & $-5.50\cdot 10^{-4}$ & $-5.39\cdot 10^{-4}$ \\ \hline
\end{tabular} \bigskip

Table 4. Effective pairing interaction in the two-dimensional isotropic model.
\end{center}

One finds, that the contributions decay approximately proportional to
$T$. The same tendency can be seen for the $f$- and $g$-wave. This is in
agreement with the prediction by Feldman et al.\cite{Feldman}, that at
zero-temperature in second order of the coupling only the s-wave part of the
pair-interaction is changed, but no contribution to higher angular momenta is
obtained.

\section{HUBBARD MODEL}

Explicit calculations were performed for the Hubbard model on a square
lattice described by the Hamiltonian
\be
H=-t\sum_{\langle r',r\rangle,s} c\dg_{r's} c_{rs} 
+ U \sum_r (n_{r\uparrow}-\frac 12) (n_{r\downarrow}-\frac 12),
\ee
where the hopping-terms are summed over nearest neighbors.

\subsection{Symmetries}

As long as $\eps_{q_0-k}=-\eps_k$ we find a number of identities. Let us begin
with
\be
W_{\rm A} = -\frac{U^2}{\Omega} \sum_{p_1,p_2} 
\frac{(1-n_{p_1}-n_{p_2})(\eps_{p_1}+\eps_{p_2})}
{(\eps_{p_1}+\eps_{p_2})^2+(\eps_k-\eps_q)^2}
\delta_{p_1+p_2,k+q+q_0}
\ee
If one replaces under the sum $p_1$ by $p_1+q_0$ and $p_2$ by $p_2-q_0$ and
considers only the 1 in the first paranthesis, one observes that this
contribution changes into its negative. Therefore we may skip the 1. Moreover
the expression is symmetric with respect to the exchange of $p_1$ and $p_2$.
Thus we may write
\be
W_{\rm A} = \frac{2U^2}{\Omega} \sum_{p_1,p_2} 
\frac{n_{p_1}(\eps_{p_1}+\eps_{p_2})}
{(\eps_{p_1}+\eps_{p_2})^2+(\eps_k-\eps_q)^2}
\delta_{p_1+p_2,k+q+q_0}.
\ee
If we now replace $p_2$ by $p_2+q_0$, then we obtain
\be
W_{\rm A} = \frac{2U^2}{\Omega} \sum_{p_1,p_2} 
\frac{n_{p_1}(\eps_{p_1}-\eps_{p_2})}
{(\eps_{p_1}-\eps_{p_2})^2+(\eps_k-\eps_q)^2}
\delta_{p_1+p_2,k+q}.
\ee
Similarly the expressions for $W_{\rm B}$ and $W_{\rm C}$ can be rewritten
\bea
W_{\rm B} = -\frac{2U^2}{\Omega} \sum_{p_1,p_2} 
\frac{n_{p_1}(\eps_{p_1}-\eps_{p_2})}
{(\eps_{p_1}-\eps_{p_2})^2+(\eps_k-\eps_q)^2}
\delta_{p_2-p_1,k+q}, \\
W_{\rm C} = -\frac{2U^2}{\Omega} \sum_{p_1,p_2} 
\frac{n_{p_1}(\eps_{p_1}-\eps_{p_2})}
{(\eps_{p_1}-\eps_{p_2})^2+(\eps_k-\eps_q)^2}
\delta_{p_2-p_1,k-q}.
\eea
From these expressions it is apparent, that
\be
W_{\rm B}(k,q)=W_{\rm C}(k,-q)=-W_{\rm A}(k,q) \label{BCA}
\ee

A second similar relation is obtained for $W_{\rm H}$ and $W_{\rm Y}$. In
these cases $\eps^a=\eps^b$ and
\be
W_{\rm H}(k,q)=W_{\rm Y}(k,q_0-q)
\ee
hold.
If moreover $\mu=0$ (half-filling), then by replacing $p_1$ to $q_0-p_1$ one
obtains
\be
W_{\rm F}(k,q)=-W_{\rm Y}(k,q).
\ee

Another class of symmetries is the following
\be
W_{\rm A,C,Y}(k,q)=W_{\rm A,C,Y}(k+q_0,q+q_0). \label{q0}
\ee
In the first two cases $\eps^a$ and $\eps^b$ interchange. Thus the
equation holds. In the last case $\eps^a$ and $-\eps^b$ have to be
exchanged. If we simultaneously exchange the momenta $p_1$ and $-p_2$ in
(\ref{BHCY}), then also the last equality is obtained.

\subsection{Free Energy in Molecular Field Approximation}

After having calculated the effective interaction we are left with a problem
which is much simpler then the original one. The reason is that the original
two-particle interaction is a function of three independent momenta. Here,
however, the interaction depends only on two independent momenta. Thus the
fluctuations, which make calculations for the original interaction difficult
are basically eliminated. It is now sufficient to find the minimum of the free
energy within a Hartree-Fock-Bogoliubov approximation. In practice we will
start from the symmetric state and investigate, whether this state is stable
against fluctuations of the order-parameters. Thus we have to expand the free
energy
\be
F=\langle H \rangle -T S,
\ee
in the fluctuations around the symmetric state ($S$ denotes the entropy). 
For this purpose we represent $\langle H \rangle $ as a function of the
expectations
\bea
\langle a\dg_k a\dg_l \rangle = \Delta_{kl},
\quad \Delta_{kl}=-\Delta_{lk},
\quad \langle a_l a_k \rangle = \Delta^*_{kl} \\
\langle a\dg_k a_l \rangle = n^0_k \delta_{kl} + \nu_{kl},
\quad \nu_{kl}=\nu^*_{lk}.
\eea
We assume $\Delta$ and $\nu$ to be small quantities and expand the entropy up
to second order in these quantities. After some calculation one obtains
\bea
S&=&S^0 +k_{\rm B} \beta \sum_k (\eps^0_k-\mu) \nu_{kk} \nn
&-&\frac{k_{\rm B}}2 \sum_{kk'} |\nu_{kk'}|^2 
f(\beta(\eps^0_k-\mu),\beta(\eps^0_{k'}-\mu)) \nn
&-&\frac{k_{\rm B}}2 \sum_{kk'} |\Delta_{kk'}|^2
f(\beta(\eps^0_k-\mu),\beta(\mu-\eps^0_{k'})),
\eea
with
\bea
S^0&=&-k_{\rm B} \sum_k \left( n^0_k \ln n^0_k + (1-n^0_k) \ln(1-n^0_k)\right),
\\ f(x,y)&=&\frac{x-y}{\frac 1{\ex{y}+1}-\frac 1{\ex{x}+1}}
=\frac{x-y}{\ex{x}-\ex{y}}(\ex{x}+1)(\ex{y}+1).
\eea
In total we have to consider four channels, two particle-hole and two
particle-particle-channels, since the total momentum can be 0 and $q_0$.
In all cases we have to distinguish between singlet- and triplet-excitations.

\subsubsection{Particle-particle channel, $q=0$, interaction $V_{\rm B}$}

In this channel we consider the expectation values
\bea
\langle c\dg_{k,s_1} c\dg_{-k,s_2} \rangle &=& \eps_{s_1,s_2} \Delta_k^{\rm s*} 
+ (\eps \sigma^{\alpha})_{s_1,s_2} \Delta_k^{\rm t\alpha*} \\
\langle c_{-k,s_2} c_{k,s_1} \rangle &=& \eps_{s_2,s_1} \Delta_k^{\rm s}
+(\sigma^{\alpha} \eps)_{s_2,s_1} \Delta_k^{\rm t\alpha}
\eea
with $\eps=\sigma^y$. They obey the symmetries
\be
\Delta_k^{\rm s} = \Delta_{-k}^{\rm s}, \quad 
\Delta_k^{\rm t\alpha} = -\Delta_{-k}^{\rm t\alpha}.
\ee
Due to these symmetries $s$- and $d$-wave pairing is of singlet type,
$p$-wave pairing of triplet type. The interaction reads
\be
H_{\rm B} = \frac 1{2\Omega} \sum_{k,q,s,s'} V_{\rm B}(k,q) 
 c\dg_{ks} c_{qs} c\dg_{-ks'} c_{-qs'}, \label{HB}
\ee
which yields the energy
\bea
E_{\rm B} &=& \frac 1{2\Omega} \sum_{k,q,s,s'} V_{\rm B}(k,q)
\big(\eps_{ss'} \Delta^{\rm s*}_k
+ (\eps\sigma^{\alpha})_{ss'} \Delta^{t\alpha*}_k \big)
\big(\eps_{s's} \Delta^{\rm s}_q
+ (\sigma^{\beta}\eps)_{s's} \Delta^{\rm t\beta}_q \big) \nn
&=& \frac 1{\Omega} \sum_{k,q} V_{\rm B}(k,q) 
\big( \Delta_k^{\rm s*} \Delta_q^{\rm s}
+\Delta_k^{\rm t\alpha*} \Delta_q^{\rm t\alpha} \big).
\eea
and the entropy
\bea
S_{\rm B}&=&-k_{\rm B} \sum_k f_{\rm B}(k) \big( \Delta_k^{\rm s*}
\Delta_k^{\rm s} +\Delta_k^{\rm t\alpha*} \Delta_k^{\rm t\alpha} \big), \\
f_{\rm B}(k) &=& f(\beta(\eps_k-\mu),\beta(\mu-\eps_k)).
\eea

\subsubsection{Particle-hole channel, $q=0$, interactions $V_{\rm H}$ and
$V_{\rm F}$}

In this channel the expectation values
\be
\langle c\dg_{ks} c_{ks'} \rangle = \delta_{s,s'} (n_k^0+\nu^{\rm s}_k)
+ \sigma^{\alpha}_{s,s'} \nu^{\rm t\alpha}_k
\ee
are considered. They obey the symmetries
\be
\nu^{\rm s*}_k = \nu^{\rm s}_k, \quad 
\nu^{\rm t\alpha*}_k = \nu^{\rm t\alpha}_k
\ee

The corresponding part of the interaction reads
\bea
H_{\rm HF}&=&\frac 1{2\Omega} \sum_{k,q,s,s'} 
V_{\rm H}(k,q) :c\dg_{ks} c_{ks} c\dg_{qs'} c_{qs'}: \nn
&+&\frac 1{2\Omega} \sum_{k,q,s,s'} V_{\rm F}(k,q) 
:c\dg_{ks} c_{qs} c\dg_{qs'} c_{ks'}:. \label{HHF}
\eea
Its contribution to the energy is
\bea
E_{\rm HF} &=& \frac 2{\Omega} \sum_{k,q} V_{\rm H}(k,q) \nu^{\rm s}_k \nu^{\rm
s}_q \nn &-&\frac 1{2\Omega} \sum_{k,q,s,s'} V_{\rm F}(k,q) 
\big(\delta_{s,s'} \nu^{\rm s}_k 
+ \sigma^{\alpha}_{s,s'} \nu^{\rm t\alpha}_k \big)
\big(\delta_{s',s} \nu^{\rm s}_q 
+ \sigma^{\beta}_{s',s} \nu^{\rm t\beta}_q \big) \nn
&=& \frac 1{\Omega} \sum_{k,q} (2V_{\rm H}(k,q)-V_{\rm F}(k,q)) 
\nu^{\rm s}_k \nu^{\rm s}_q 
-\frac 1{\Omega} \sum_{k,q} V_{\rm F}(k,q) \nu^{\rm t\alpha}_k \nu^{\rm
t\alpha}_q,
\eea
and its entropy reads
\bea
S_{\rm HF}&=&-k_{\rm B} \sum_k f_{\rm H}(k) \big( (\nu_k^{\rm s})^2 
+(\nu_k^{\rm t\alpha})^2 \big) \\ 
f_{\rm H}(k) &=& f(\beta(\eps_k-\mu),\beta(\eps_k-\mu)).
\eea

\subsubsection{Particle-particle channel, momentum $q_0$, 
interaction $V_{\rm Y}$}

In this channel the expectation values
\bea
\langle c\dg_{ks} c\dg_{q_0-k,s'} \rangle &=& 
 \eps_{ss'} \Delta^{\rm s*}_k 
+(\eps \sigma^{\alpha})_{ss'} \Delta^{\rm t\alpha*}_k \\
\langle c_{q_0-ks'} c_{ks} \rangle &=& \eps_{s's} \Delta^{\rm s}_k 
+ (\sigma^{\alpha} \eps)_{s's} \Delta^{\rm t\alpha}_k
\eea
are considered with the symmetries
\be
\Delta^{\rm s}_{q_0-k}= \Delta^{\rm s}_k, \quad 
\Delta^{\rm t\alpha}_{q_0-k} = - \Delta^{\rm t\alpha}_k.
\ee

The interaction reads
\be
H_{\rm Y}=\frac 1{2\Omega} \sum_{k,q,s,s'} V_{\rm Y}(k,q)
c\dg_{ks} c_{qs} c\dg_{q_0-ks'} c_{q_0-qs'}, \label{HY}
\ee
and its contribution to the energy
\bea
E_{\rm Y} &=&\frac 1{2\Omega} \sum_{k,q,s,s'} V_{\rm Y}(k,q) 
\big( \eps_{ss'} \Delta^{\rm s*}_k 
+(\eps \sigma^{\alpha})_{ss'} \Delta^{\rm t\alpha*}_k \big)
\big(\eps_{s's} \Delta^{\rm s}_q 
+ (\sigma^{\beta}\eps)_{s's} \Delta^{\rm t\beta}_q \big) \nn
&=& \frac 1{\Omega} \sum_{k,q} V_{\rm Y}(k,q)
\big( \Delta^{\rm s*}_k \Delta^{\rm s}_q 
+ \Delta^{\rm t\alpha*}_k \Delta^{\rm t\alpha}_q \big).
\eea

Let us now divide the Brillouin zone into two halves. The contributions to
the sum with $|k|<|q_0-k|$ are denoted by a prime $'$. Then the energy reads
by use of eq. (\ref{q0})
\bea
E_{\rm Y} &=& \frac 2{\Omega} \sum^{\prime}_{k,q} V_{\rm Y}(k,q)
\big( \Delta^{\rm s*}_k \Delta^{\rm s}_q 
+ \Delta^{\rm t\alpha*}_k \Delta^{\rm t\alpha}_q \big) \nn
&&+ \frac 2{\Omega} \sum^{\prime}_{k,q} V_{\rm Y}(k,q_0-q)
\big( \Delta^{\rm s*}_k \Delta^{\rm s}_q 
- \Delta^{\rm t\alpha*}_k \Delta^{\rm t\alpha}_q \big) \nn
&=& \frac 2{\Omega} \big( \sum^{\prime}_{k,q} V_{\rm Y}(k,q)
+V_{\rm Y}(k,q_0-q) \big) \Delta^{\rm s*}_k \Delta^{\rm s}_q \nn
&&+ \frac 2{\Omega} \sum^{\prime}_{k,q} \big( V_{\rm Y}(k,q)
-V_{\rm Y}(k,q_0-q) \big)
\Delta^{\rm t\alpha*}_k \Delta^{\rm t\alpha}_q,
\eea
and the corresponding entropy
\bea
S_{\rm Y}&=& -\frac{k_{\rm B}}2 \sum_{k,s,s'} f_{\rm Y}(k)
\big( \eps_{ss'} \Delta^{\rm s*}_k 
+(\eps \sigma^{\alpha})_{ss'} \Delta^{\rm t\alpha*}_k \big)
(\eps_{s's} \Delta^{\rm s}_k 
+ (\sigma^{\beta}\eps)_{s's} \Delta^{\rm t\beta}_q \big) \nn
&=& -k_{\rm B} \sum_k f_{\rm Y}(k)
\big( \Delta^{\rm s*}_k \Delta^{\rm s}_k 
+ \Delta^{\rm t\alpha*}_k \Delta^{\rm t\alpha}_k \big) \nn
&=& -2k_{\rm B} \sum^{\prime}_k f_{\rm Y}(k)
\big( \Delta^{\rm s*}_k \Delta^{\rm s}_k 
+ \Delta^{\rm t\alpha*}_k \Delta^{\rm t\alpha}_k \big), \\
f_{\rm Y}(k) &=& f(\beta(\eps_k-\mu),\beta(\mu-\eps_{q_0+k}).
\eea

\subsubsection{Particle-hole channel, momentum $q_0$, interactions $V_{\rm A}$
and $V_{\rm C}$}

Here the expectation values
\be
\langle c\dg_{ks} c_{k+q_0s'} \rangle = \delta_{ss'} \nu^{\rm s}_k
+\sigma^{\alpha}_{ss'} \nu^{\rm t\alpha}_k
\ee
are considered, which obey the symmetries
\be
\nu^{\rm s}_{k+q_0} = \nu^{\rm s*}_k \quad
\nu^{\rm t\alpha}_{k+q_0} = \nu^{\rm t\alpha*}_k.
\ee

The interaction reads
\bea
H_{\rm AC} &=& \frac 1{2\Omega} \sum_{k,q,s,s'} V_{\rm A}(k,q)
c\dg_{ks} c_{qs} c\dg_{q+q_0s'} c_{k+q_0s'} \nn
&+&\frac 1{2\Omega} \sum_{k,q,s,s'} V_{\rm C}(k,q)
c\dg_{ks} c_{k+q_0s} c\dg_{q+q_0s'} c_{qs'}. \label{HAC}
\eea
Its contribution to the energy is
\bea
E_{\rm AC} &=& - \frac 1{2\Omega} \sum_{k,q,s,s'} V_{\rm A}(k,q)
\big( \delta_{ss'} \nu^{\rm s}_k 
+\sigma^{\alpha}_{ss'} \nu^{\rm t\alpha}_k \big)
\big( \delta_{s's} \nu^{\rm s*}_q 
+\sigma^{\alpha}_{s's} \nu^{\rm t\alpha*}_q \big) \nn
&+& \frac 2{\Omega} \sum_{k,q,s,s*} V_{\rm C}(k,q)
\nu^{\rm s}_k \nu^{\rm s*}_q \\
&=& \frac 1{\Omega} \sum_{k,q}
\big( 2V_{\rm C}(k,q) - V_{\rm A}(k,q) \big) 
\nu^{\rm s}_k \nu^{\rm s*}_q 
- \frac 1{\Omega} \sum_{k,q} V_{\rm A}(k,q)
\nu^{\rm t\alpha}_k \nu^{\rm t\alpha*}_q. \nonumber
\eea

Summation over half of the Brillouin zone yields with eq. (\ref{q0})
\bea
E_{\rm AC} &=& \frac 1{\Omega} \sum^{\prime}_{k,q}
\big( 2V_{\rm C}(k,q) - V_{\rm A}(k,q) \big)
\big( \nu^{\rm s}_k \nu^{\rm s*}_q 
+ \nu^{\rm s*}_k \nu^{\rm s}_q \big) \nn
&+& \frac 1{\Omega} \sum^{\prime}_{k,q}
\big( 2V_{\rm C}(k,q+q_0) 
- V_{\rm A}(k,q+q_0) \big)
\big( \nu^{\rm s}_k \nu^{\rm s}_q 
+ \nu^{\rm s*}_k \nu^{\rm s*}_q \big) \nn
&-& \frac 1{\Omega} \sum^{\prime}_{k,q}
V_{\rm A}(k,q)
\big(\nu^{\rm t\alpha}_k \nu^{\rm t\alpha*}_q 
+ \nu^{\rm t\alpha*}_k \nu^{\rm t\alpha}_q \big) \nn
&-&\frac 1{\Omega} \sum^{\prime}_{k,q}
V_{\rm A}(k,q+q_0)
\big( \nu^{\rm t\alpha}_k \nu^{\rm t\alpha}_q 
+ \nu^{\rm t\alpha*}_k \nu^{\rm t\alpha*}_q \big).
\eea

We now separate the real and the imaginary part of $\nu$
\bea
\nu_k = \frac 1{\sqrt2} (\bar\nu_k+i\bar{\bar\nu}_k), & &
\nu_{k+q_0} = \frac 1{\sqrt2} (\bar\nu_k-i\bar{\bar\nu}_k) \\
\nu_k\nu_q+\nu^*_k\nu^*_q &=& \bar\nu_k \bar\nu_q 
- \bar{\bar\nu}_k \bar{\bar\nu}_q \\
\nu_k\nu^*_q+\nu^*_k\nu_q &=& \bar\nu_k \bar\nu_q 
+ \bar{\bar\nu}_k \bar{\bar\nu}_q.
\eea
Then the energy reads
\bea
E_{\rm AC} &=& \frac 1{\Omega} \sum^{\prime}_{k,q}
\big( 2V_{\rm C}(k,q) - V_{\rm A}(k,q) 
+ 2V_{\rm C}(k,q+q_0) - V_{\rm A}(k,q+q_0) \big)
\bar\nu^{\rm s}_k \bar\nu^{\rm s}_q \nn
&+& \frac 1{\Omega} \sum^{\prime}_{k,q}
\big( 2V_{\rm C}(k,q) - V_{\rm A}(k,q) 
- 2V_{\rm C}(k,q+q_0) + V_{\rm A}(k,q+q_0) \big)
\bar{\bar\nu}^{\rm s}_k \bar{\bar\nu}^{\rm s}_q \nn
&-& \frac 1{\Omega} \sum^{\prime}_{k,q}
\big( V_{\rm A}(k,q) + V_{\rm A}(k,q+q_0) \big)
\bar\nu^{\rm t\alpha}_k \bar\nu^{\rm t\alpha}_q \nn
&-& \frac 1{\Omega} \sum^{\prime}_{k,q}
\big( V_{\rm A}(k,q) - V_{\rm A}(k,q+q_0) \big)
\bar{\bar\nu}^{\rm t\alpha}_k \bar{\bar\nu}^{\rm t\alpha}_q. \label{EAC}
\eea
The corresponding entropy is
\bea
S_{\rm AC} &=& - \frac{k_{\rm B}}2 \sum_{k,s,s'} f_{\rm C}(k)
\big(\delta_{ss'}\nu^{\rm s}_k 
+\sigma^{\alpha}_{ss'}\nu^{\rm t\alpha}_k \big)
\big(\delta_{s's}\nu^{\rm s*}_k 
+\sigma^{\alpha}_{s's}\nu^{\rm t\alpha*}_k \big) \nn
&=& -k_{\rm B} \sum_k f_{\rm C}(k)
\big( \nu^{\rm s}_k \nu^{\rm s*}_k
+\nu^{\rm t\alpha}_k \nu^{\rm t\alpha*}_k \big) \nn
&=& -k_{\rm B} \sum^{\prime}_k f_{\rm C}(k)
\big( (\bar\nu^{\rm s}_k)^2 +(\bar{\bar\nu}^{\rm s}_k)^2
+ (\bar\nu^{\rm t\alpha}_k)^2 +(\bar{\bar\nu}^{\rm t\alpha}_k)^2 \big) \nn
f_{\rm C}(k) &=& f(\beta(\eps_k-\mu),\beta(\eps_{k+q_0}-\mu)).
\eea

\subsection{Variation of the Free Energy}

The expression for the free energy which is bilinear in $\nu$ and $\Delta$,
resp., has to be checked with respect to its stability. That is, as soon as
some $\nu$ or $\Delta$ different from zero yields a lower free energy than for
the symmetric state for which all $\nu$ and $\Delta$ vanish, then the
symmetric state is unstable and the system will approach a symmetry broken
state. This is the indication for a phase transition. The expression for the
free energy has the form
\bea
\beta F &=& \frac 1{\Omega} \sum_{k,q} (\beta U)(1+\frac Ut V_{k,q})
\Delta_k^* \Delta_q + \sum_k f_k \Delta_k^* \Delta_k \\
&=& \sum_{k,q} \left( \frac Ut A_{k,q}+ (\frac Ut)^2 B_{k,q}
+\delta_{k,q} \right) \, \sqrt{f_k}\Delta_k^* \, \sqrt{f_q}\Delta_q
\eea
with
\be
A_{k,q}=\frac{\beta t}{\Omega \sqrt{f_k f_q}}, \quad
B_{k,q}=\frac{\beta t V_{k,q}}{\Omega \sqrt{f_k f_q}}.
\ee
A similar bilinear contribution is obtained for $\nu$ instead of $\Delta$,
which is handled in the same way.
Here the factor $U^2/t$ is extracted from the matrix elements $V_{k,q}$.
These matrix elements and the entropy coefficients $f_k$ depend on $\beta t$
and $\beta \mu$. The same is true for the coefficients $A_{k,q}$ and $B_{k,q}$.

The calculation is performed for the various representations under the group 
$C_4=4mm$. The representations of the even-parity states are
one-di\-men\-sio\-nal. We denote them by $s_+=s_1$,
$s_-=g=s_{xy(x^2-y^2)}$, $d_+=d_{x^2-y^2}$, $d_-=d_{xy}$.
The odd-parity representation is two-dimensional, here simply denoted by $p$.
Moreover in the channels $V_{\rm A}$, $V_{\rm C}$, and $V_{\rm Y}$ we can
distinguish between eigen solutions $\nu_k=\pm\nu_{k+q_0}$, and 
$\Delta_k=\pm\Delta_{k+q_0}$, resp. In total the calculation is performed for
45 channels.

For each channel the eigenvalues $\lambda$ of $\frac Ut A + (\frac Ut)^2 B$
have to be determined. Whenever the lowest eigenvalue (i.e. the most negative)
equals $-1$, then a critical $(U/t)_{\rm c}$ is reached.
For the $s_+$ representation (in case of $V_{\rm A}$, $V_{\rm C}$, $V_{\rm Y}$
only for $\nu_k=+\nu_{k+q_0}$, $\Delta_k=+\Delta_{k+q_0}$) one has to find the
solution by iterating the eigenvalue equation as a function of $U/t$. In the
case of all other representations the $A$-term does not contribute. Therefore
then we determine the lowest eigenvalue $\lambda$ of $B$, and obtain 
$(U/t)_{\rm c}=1/\sqrt{-\lambda}$.

\section{NUMERICAL RESULTS}

\subsection{Numerics}

In a first step we determine the matrix elements as given above on a grid of 
$2n_{\rm h} \times 2n_{\rm h}$ lattice points in the Brillouin zone. The
results\cite{Grote} presented below are calculated for $n_{\rm h}=16$. The
calculation of the matrix-elements takes most of the computer time. It
increases with the sixth power of $n_{\rm h}$.

It is obvious that in a number of cases the denominator becomes extremely
small or even vanishes. We remember, that the sums (\ref{sum}) are of the form
\be
\sum_p \frac{b_p z_p}{z_p^2+e^2}, \label{bze}
\ee
where $e$ is independent of $p$. In particular if $e$ vanishes then we have
basically a main value integral
\be
\int \de^2 p \frac{b_p}{z_p}
\ee
where $z_p$ can vanish. In order to avoid, that we sum up terms close to a
zero of $z_p$, which would yield eratic contributions we proceed as follows:
We average over several points in momentum-space in the vicinity of
$p$. More precisely if we have to sum
\be
\sum_{p_1,p_2} \frac{z_{k,q,p_1,p_2}}{n_{k,q,p_1,p_2}} \delta_{k+q,p_1+p_2},
\ee
then we calculate $z$ and $n$ not only for the given $(k,q,p_1,p_2)$, but also
for $(k+\delta,q,p_1+\delta,p_2)$, $(k+\delta,q,p_1,p_2+\delta)$,
$(k,q+\delta,p_1+\delta,p_2)$, $(k,q+\delta,p_1,p_2+\delta)$,
$(k+\delta,q-\delta,p_1,p_2)$, $(k,q,p_1+\delta,p_2-\delta)$ for four different
$\delta$s. They are $\delta=(\pm \frac{\delta p}2,0)$ and  $\delta=(0,\pm
\frac{\delta p}2)$, where $\delta p$ is the momentum-spacing for the evaluation
of the sums. Then we attribute to these the weights $g n$ and to
$(k,q,p_1,p_2)$ the corresponding weight $n$, which means that we use
\be
\sum_{p_1,p_2} \frac{z_{k,q,p_1,p_2}+g\sum_{\{\delta\}} z_{..\pm\delta..}}
{n_{k,q,p_1,p_2}+g\sum_{\{\delta\}} n_{..\pm\delta..}} \delta_{k+q,p_1+p_2}
\label{num}
\ee
with some choice of $g$. Since $n$ is never negative we obtain nearly always a
positive denominator. An exception are the $V_{\rm H}$ and 
$V_{\rm F}$-terms which obey $k=q=p$
and $2k=q_0$ modulo reciprocal lattice vector, and the $V_{\rm Y}$-terms with
$p=q-q_0=-k$ and $2k=q_0$ (modulo reciprocal lattice vector). In these cases
all $n$s vanish and we replace the term by the average of the four terms at 
$p\pm(\delta p,0)$ and $p\pm(0,\delta p)$.

Since the matrix-elements are calculated by summing over a grid in the
Brillouin zone, the question arises, how reliable are the results? Obviously
at low temperatures the occupation numbers $n$ vary rapidly. So the result may
depend on the points in our grid. There are basically two criteria: \\
(i) In the channels $V_{\rm 2F}$, $V_{\rm 2H}$, and $V_{\rm 2Y}$ the
denominator in (\ref{bze}) contains an $e$ which vanishes identically.
Despite of the procedure described above one has to assume, that the calculated
matrix-elements are less precise than those for $V_{\rm B}$, $V_{\rm A}$, and
$V_{\rm C}$. \\
(ii) The increase of the entropy factors $f$ determines the extension in the
wave-vector space contributing essentially to the order-parameter. Denoting
$x=\beta(\eps-\mu)$ one observes, that $f_{\rm B}=f(x,-x)$ increases like
$2|x|$, whereas $f_{\rm H}=f(x,x)$ increases like $\ex{|x|}$ for large $|x|$.
Thus the contributing phase space for $V_{\rm B}$ is larger than for 
$V_{\rm H,F}$. The other interactions lie in between. Indeed it turns out,
that the results for the particle-hole channels coming from $V_{\rm H,F}$ show
deviations between different $n_{\rm h}$s at low temperatures, that is below
$0.05t$.

\subsection{Results}

In the following figures we show the critical values of the lowest
lying $(U/t)_{\rm c}$ as function of $T/t$ for different values of $\mu/t$.
We have chosen $g=0.01$ in (\ref{num}). Calculations for $g=1/24$ instead
yield practically no difference. 

In the following table we summarize the frequently observed low-lying
instabilities. In the second column we charactarize the symmetries: $ph$
stands for particle-hole, $pp$ for particle-particle, $si$ for singlet, 
$tr$ for triplet, $q_{\pm}$ for $\nu_{k+q_0}=\pm \nu_k$, and 
$\Delta_{k+q_0}=\pm \Delta_k$, otherwise 0. If there are instabilities in
another channel with an $(U/t)_{\rm c}$ less than that for the
Pomeranchuk-instability with ($s_-=g$) and less than 10, then the lowest one
of those is also shown.

\begin{center}
\pick{\epsfig{file=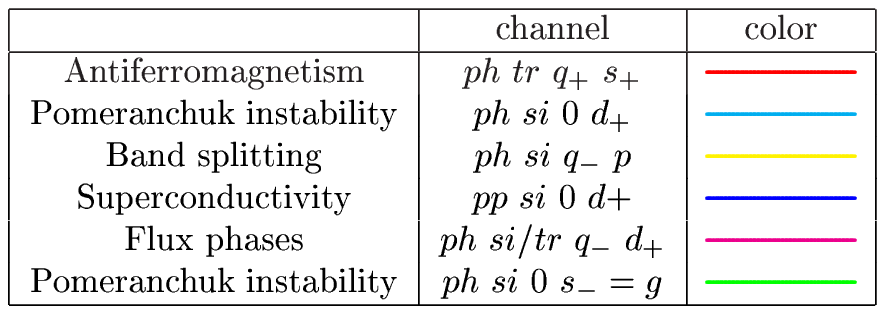,scale=1,angle=0}}
{\epsfig{file=huwtab.ps,scale=1,angle=0}} \bigskip

Table 5. Channels with low-lying $(U/t)_{\rm c}$.
\end{center}
\medskip

\begin{figure}
\pick{\epsfig{file=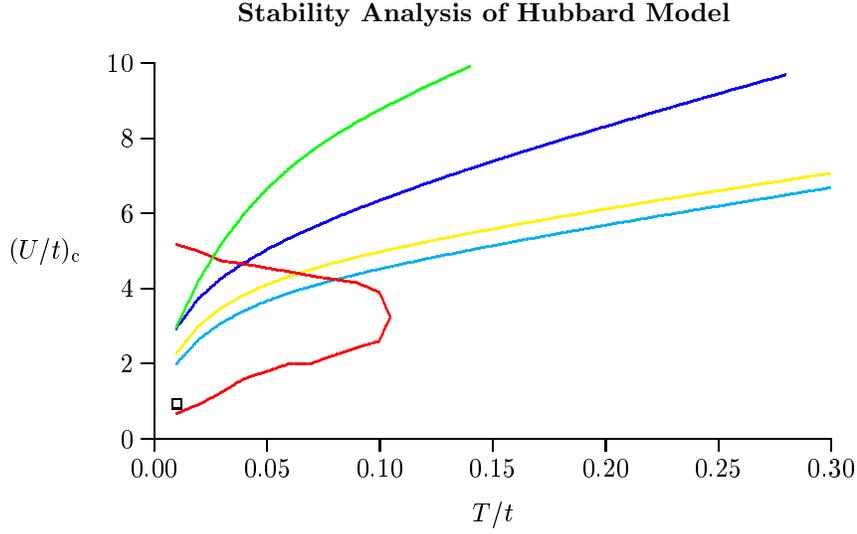,scale=1,angle=0}}
{\epsfig{file=huw1611.ps,scale=1,angle=0}} \\
\caption
{$(U/t)_{\rm c}$ at $\mu/t=0$. Doping $x=0$.
Also shown: At $T/t=$ 0.01 symmetry ($ph$ $tr$ $s_+$) (ferromagnetism,
disappears for larger lattices).}
\label{fig1}
\end{figure}

\begin{figure}
\pick{\epsfig{file=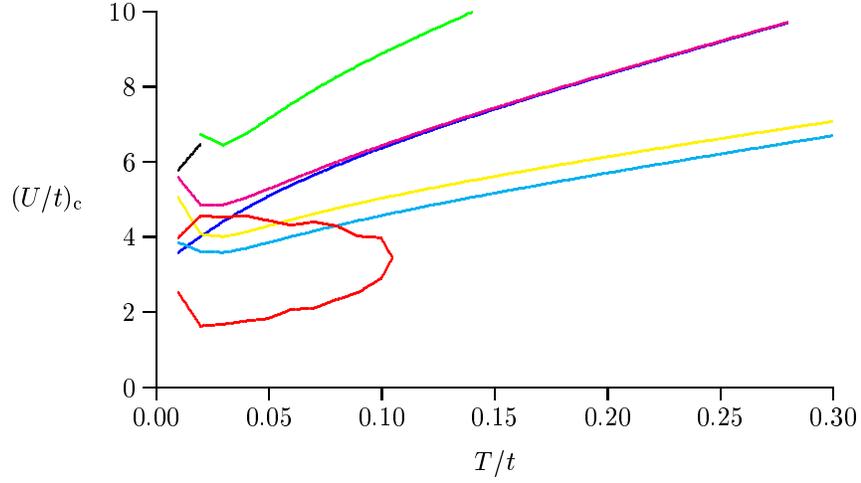,scale=1,angle=0}}
{\epsfig{file=huw1612.ps,scale=1,angle=0}} \\
\caption
{$(U/t)_{\rm c}$ at $\mu/t=1/24$. 
Doping varies from $x=0.0625$ to $x=0.0174$.
Also shown:
For $T/t= 0.01$ to $0.02$ with symmetry ($pp$ $tr$ $q-$ $p$).}
\label{fig2}
\end{figure}

\begin{figure}
\pick{\epsfig{file=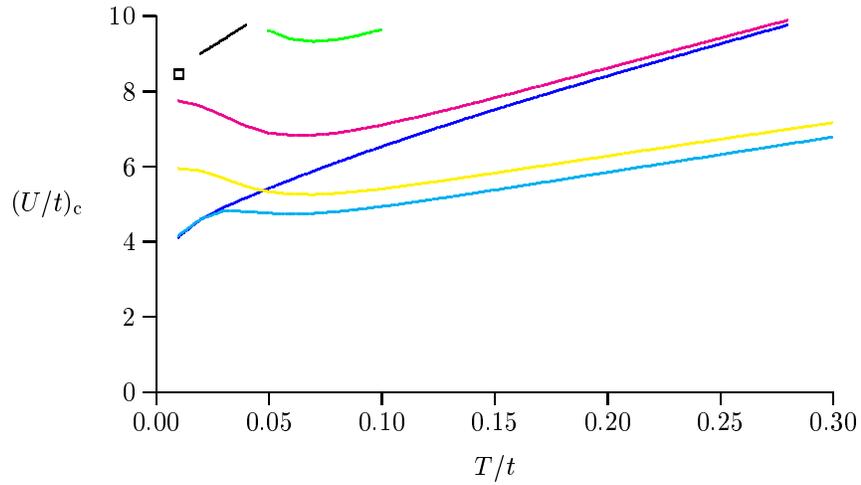,scale=1,angle=0}}
{\epsfig{file=huw1614.ps,scale=1,angle=0}} \\
\caption
{$(U/t)_{\rm c}$ at $\mu/t=1/8$.
Doping varies from $x=0.0807$ to $x=0.0519$.
Also shown:
At $T/t= 0.01$ symmetry ($ph$ $tr$ $p$).
From $T/t= 0.02$ to $0.04$ with symmetry ($pp$ $si$ $s_-=g$).}
\label{fig3}
\end{figure}

\begin{figure}
\pick{\epsfig{file=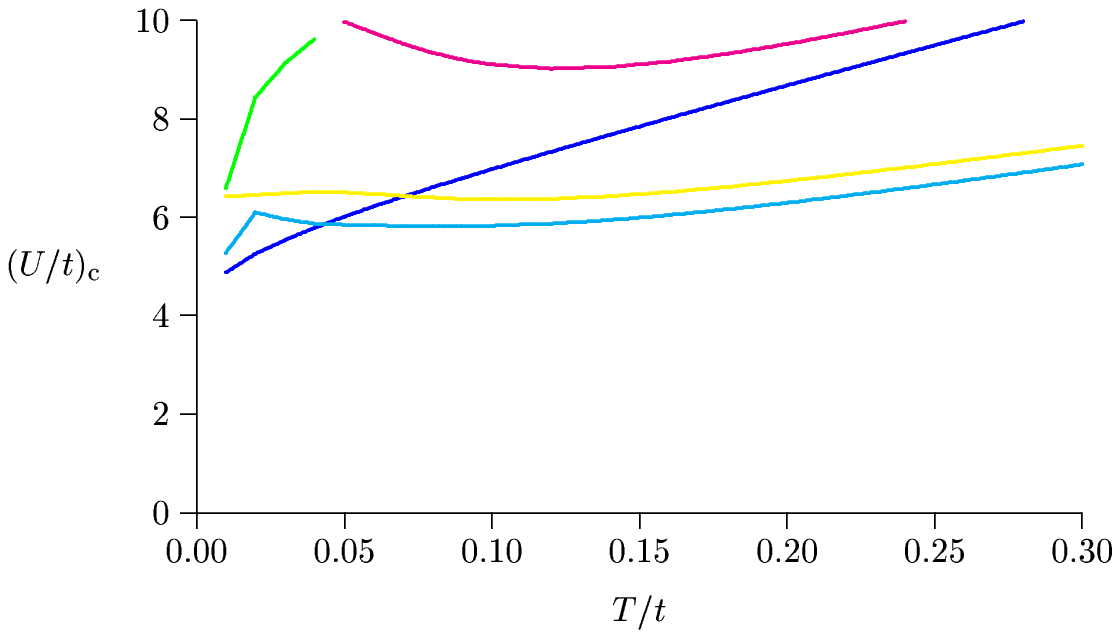,scale=1,angle=0}}
{\epsfig{file=huw1616.ps,scale=1,angle=0}} \\
\caption
{$(U/t)_{\rm c}$ at $\mu/t=1/4$.
Doping varies from $x=0.1159$ to $x=0.1030$.}
\label{fig4}
\end{figure}

As expected at half-filling ($\mu/t=0$) the lowest lying is the
an\-ti\-fer\-ro\-mag\-ne\-tic one. It turns out, that the second order
contribution to the antiferromagnetic channel suppresses antiferromagnetism
(at least in the $s$-channel). Therefore antiferromagnetism disappears at
larger values of $U/t$. This is expected since in the strong-coupling limit the
magnetic interaction vanishes with $J=4t^2/U$ in the $t$-$J$-model.
As we leave half-filling the antiferromagnetism becomes weaker and disappears
at some doping. This observation is remarkable, since although we work with a
weak-coupling calculation, we obtain results expected at strong couplings. 

The next instability is a Pomeranchuk-instability. It has recently been
observed from RG-calculations by Halboth and Metzner\cite{MetznerII}. It
should be mentioned, that our calculations become less reliable for the 
$ph 0$-sector at very low temperatures. Thus we do not know, whether (except
from the antiferromagnetic instability) it is the lowest lying one as $T$
approaches zero. Depending on the choice of  $n_{\rm h}$ the slope of the
curve remains positive at low temperature or it becomes negative.

The next instability is a particle-hole instability of singlet type with
$q_-$ and $p$-type symmetry. It corresponds to an strengthening and weakening
of the hopping matrix-elements at alterning bonds along the $x$- or
$y$-direction or both superimposed. This instability leads to a splitting into
two bands. Within the present approximation they are not separated by a gap.
One may speculate, that in higher orders they develop into the two Hubbard
bands. We observe, however, that at sufficiently low temperatures either
antiferromagnetism or $d$-wave superconductivity yield a lower critical
$(U/t)$.

Then the superconducting $d_{x^2-y^2}$ instability follows, which is much
stronger than superconducting instabilities of other symmetries.
\begin{center}
\begin{tabular}{|c|rr|}
\hline symmetry & $T=0.1 t$ & $T=0.03 t$ \\ \hline
$s_+$ & 12.51 \dots -0.10 & 25.90 \dots -0.39 \\
$s_-=g$ & 0.00 \dots -0.71 & 0.00 \dots -1.75 \\
$d_+$   & 0.01 \dots -2.49 & 0.01 \dots -5.49 \\
$d_-$    & 0.69 \dots -0.04 & 2.36 \dots -0.06 \\
$p$ & 1.35 \dots -0.32 & 2.74 \dots -0.97 \\
\hline
\end{tabular} \medskip

Table 6. The range of the eigenvalues $\lambda$ at half-filling\cite{WegnerII}
multiplied by 100. Here only the second-order contribution is taken into
account for $s_+$. %Compare\cite{WegnerII}
\end{center}

Finally we observe a flux-phase instability, which has been discussed by
Kotliar\cite{Kotliar}, and by Affleck and Marston\cite{Affleck}.
It has been recently discussed by Chakravarty et al. as
$d$-density wave-order\cite{Chakravarty}. The singlet and triplet eigenvalues
are degenerate for even parities (apart from ($q_+$ $s_+$)), since according to
table 1 and eqs. (\ref{BCA},\ref{EAC}) the singlet interaction 
$2V_{\rm 2C}-V_{\rm 2A} = 2 W_{\rm C}+W_{\rm A} = -W_{\rm A}$ equals the
triplet interaction $-V_{\rm 2A}$. At $\mu=0$ it is moreover degenerate with
the superconducting instability, since 
$V_{\rm 2B}=W_{\rm B}=-W_{\rm A}=-V_{\rm 2A}$ and for vanishing $\mu$ also the
entropy factors $f_{\rm B}=f_{\rm C}$ agree.

At even higher values another Pomeranchuk-instability appears with $s_-=g$
wave-character. A few other low-lying instabilities observed only at special
parameters are shown in the figures \pick{in black.}{(solid line).}
The ferromagnetic instability at $T=0.01t$ and $\mu=0$ may be a remnant of the
Nagaoka-ferromagnetism\cite{Nagaoka}, although it is expected at larger values
of $U/t$. For $n_{\rm h}=24$ it does no longer appear in our calculation even
at $T/t=0.01$.

One phase may suppress another phase. To which extend two order-parameters can
coexist with each other is another question, which should be investigated in
the future.

\subsection{Conclusion}

In these calculations we obtain many critical couplings $(U/t)_{\rm c}$ of the
order of 5. Since this is not a small number, calculations which do not handle
the coupling perturbatively, should be performed. Nevertheless it becomes
clear that this type of calculation gives a good estimate of the most
important instabilities. We have found all commonly dicussed types of
order. Moreover, a $p$-wave instability which yields band-splitting,
appears. Although our calculation is performed for weak coupling, a number of
effects normally obtained in the strong-coupling limit are reproduced
reasonably well by means of the flow equations.

\subsection*{ACKNOWLEDGMENTS}

We are indebted to Andreas Mielke and to Matthias Vojta for useful comments.

\end{document}